\documentclass[prl,aps,superscriptaddress,twocolumn,amssymb,amsmath,longbibliography]{revtex4-1}
\usepackage{graphicx}
\usepackage{theorem}
\usepackage{dcolumn}
\usepackage{bm}
\usepackage{mathrsfs}
\usepackage[cp1251]{inputenc}
\usepackage{setspace}

\begin{document}

\title{The true mechanism of spontaneous order from turbulence in two-dimensional superfluid manifolds}
%Angular Momentum Transport in Bose-Einstein Condensates by Solitonic ``Corkscrew''
%Torque and angular momentum transfer by nontopological phase defects in Bose-Einstein Condensates
\author{Toshiaki Kanai}
\affiliation{National High Magnetic Field Laboratory, 1800 East Paul Dirac Drive, Tallahassee, Florida 32310, USA}
\affiliation{Department of Physics, Florida State University, Tallahassee, Florida 32306, USA}

\author{Wei Guo}
\email[Corresponding author: ]{wguo@magnet.fsu.edu}
\affiliation{National High Magnetic Field Laboratory, 1800 East Paul Dirac Drive, Tallahassee, Florida 32310, USA}
\affiliation{Mechanical Engineering Department, FAMU-FSU College of Engineering, Florida State University, Tallahassee, Florida 32310, USA}

\date{\today}

\begin{abstract}
In a two-dimensional (2D) turbulent fluid containing point-like vortices, Lars Onsager predicted that adding energy to the fluid can lead to the formation of persistent clusters of like-signed vortices, i.e., Onsager vortex (OV) clusters. In the evolution of 2D superfluid turbulence in a uniform disk-shaped Bose-Einstein condensate (BEC), it was discovered that a pair of OV clusters with opposite signs can form without any energy input. This striking spontaneous order was explained as due to a vortex evaporative-heating mechanism, i.e., annihilations of vortex-antivortex pairs which remove the lowest-energy vortices and thereby boost the mean energy per vortex. However, in our search for exotic OV states in a boundaryless 2D spherical BEC, we found that OV clusters never form despite the annihilations of vortex pairs. Our analysis reveals that contrary to the general belief, vortex-pair annihilation emits intense sound waves, which damp the motion of all vortices and hence suppress the formation of OV clusters. We also present unequivocal evidences showing that the true mechanism underlying the observed spontaneous OV state is the escaping of vortices from the BEC boundary. Uncovering this mechanism paves the way for a comprehensive understanding of emergent vortex orders in 2D manifolds of superfluids driven far from equilibrium.
\end{abstract}
%\pacs{03.75.Lm, 03.75.Kk, 03.65.Vf}
\maketitle

In two-dimensional (2D) turbulent flows such as in soap films~\cite{Kellay-2002-RPP} and Jupiter's atmosphere~\cite{Adriani-2018-Nature}, large-scale persistent vortex structures are often observed. The appearance of these large-scale vortices can be understood in terms of a simplified point-vortex model proposed by Onsager~\cite{Onsager-1949-NCS}: when energy is continuously injected into a finite-sized 2D fluid containing many point-like vortices, the like-signed vortices must eventually aggregate to form large clusters (i.e., Onsager vortex (OV) clusters) in order to sustain the high kinetic energy of the fluid. This ordered OV state is associated with a negative temperature since it has more energy but less entropy as compared to a state with randomly distributed vortices~\cite{Onsager-1949-NCS}. While Onsager's model has provided valuable insights into 2D turbulence in general~\cite{Eyink-2006-MPR, Boffetta-2012-ARFM}, it is particularly relevant to 2D superfluids, such as planar Bose-Einstein condensates (BECs)~\cite{Johnstone-2019-Sci, Gauthier-2019-Sci} and superfluid helium films~\cite{Sachkou-2019-Sci,Varga-2020-PRL}, where the vortices are indeed point-like topological defects with a quantized circulation~\cite{Donnelly-book}.

\begin{figure}[htb]
\includegraphics[width=1.0\linewidth]{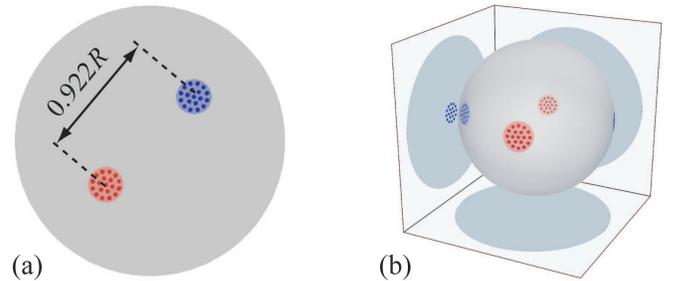}
\caption{Schematics showing the limiting configuration of OV clusters in 2D BECs with zero angular momentum in a) planar disk geometry and b) spherical shell geometry. The points of different colors represent vortices of different signs.} \label{Fig1}
\end{figure}

Surprisingly, recent numerical simulations of 2D turbulence in uniform disk-shaped BECs uncovered that a pair of OV clusters with opposite signs can form even in the absence of any energy input~\cite{Simula-2014-PRL,Billam-2014-PRL}. This intriguing spontaneous emergence of order from chaos has prompted extensive subsequent research~\cite{Yu-2016-PRA,Reeves-2017-PRL, Groszek-2018-PRL, Pakter-2018-PRL, Han-2018-JPSJ, Maestrini-2019-JSP}. A widely accepted explanation is that this emergent order is caused by a vortex evaporative-heating mechanism~\cite{Simula-2014-PRL,Billam-2014-PRL}, i.e., annihilations of vortex-antivortex pairs at close separation. Such pairs of vortices induce negligible flows in the BEC. Therefore, their annihilations merely decrease the number of vortices but retain the total energy of the vortex system, which thereby increases the mean energy per vortex. For a disk-shaped BEC with a radius $R$ carrying zero angular momentum but sufficient energy, it has been shown that as the vortices keep annihilating, the vortex system can evolve into the negative temperature state and eventually approach a limiting configuration consisting of two concentrated vortex clusters separated symmetrically around the disk center by about $0.922R$~\cite{Yu-2016-PRA}, as shown in Fig.~\ref{Fig1}~(a). This limiting configuration gives the highest kinetic energy per vortex.

\begin{figure*}[htb]
\includegraphics[width=1.0\linewidth]{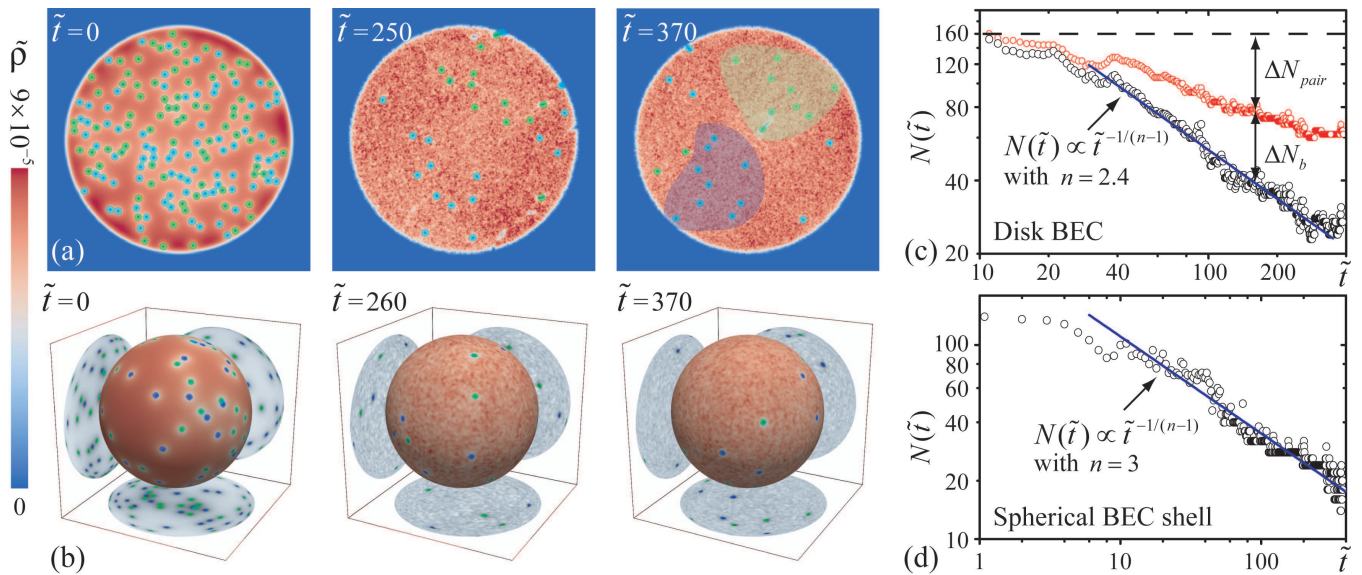}
\caption{(a) and (b) show the evolution of the condensate density $\tilde{\rho}=|\tilde{\psi}|^2$ in the GP model for the quasi-2D BEC in the disk geometry and the spherical shell geometry, respectively. The vortices and antivortices are marked with dots of different colors for better visibility. The shaded regions in the disk BEC signify the places where coherent OV clusters are seen. (c) and (d) show the evolution of the total vortex number $N(\tilde{t})$ (black circles). The red circles in the disk BEC case give the partition of the decayed vortices due to the pair-annihilation process $\Delta N_{pair}$ and due to vortices escaping the boundary $\Delta N_{b}$.}
\label{Fig2}
\end{figure*}

Recently, there have been increasing interests in BECs confined in a spherical shell geometry~\cite{Tononi-2019-PRL, Tononi-2020-PRL,Padavi-2020-PRA, Moller-2020-NJP}. Creating such a curved BEC manifold using a spherical bubble trap was proposed two decades ago~\cite{Zobay-2001-PRL}, but later research showed that this could be achieved only in microgravity~\cite{Colombe-2004-EPL,Harte-2018-PRA}. Nevertheless, this technical barrier was conquered recently due to the installation of the NASA cold atom laboratory at the international space station~\cite{Elliott-2018-npjM, Lundblad-2019-npjM}. Unlike the disk BEC case, the formation of any dipole OV-cluster configuration in 2D turbulence on a spherical surface is always associated with a finite angular momentum and therefore is prohibited if the BEC has zero angular momentum to begin with. In this situation, a novel quadrupole limiting configuration with two pairs of like-signed OV clusters across two perpendicular diameters is expected (see Fig.~\ref{Fig1}~(b)), since the corresponding flow field carries the highest kinetic energy with zero angular momentum.

%It is worthwhile noting that 2D spherical BEC is not an imaginary concept. Creating a 2D BEC shell using a spherical bubble trap was proposed two decades ago~\cite{Zobay-2001-PRL}. Later research showed that such a curved BEC manifold could be achieved only in microgravity~\cite{Colombe-2004-EPL,Harte-2018-PRA}. Recently, this technical barrier was conquered due to the installation of the NASA cold atom laboratory at the international space station~\cite{Elliott-2018-npjM, Lundblad-2019-npjM}, which has triggered many recent studies~\cite{Tononi-2019-PRL, Tononi-2020-PRL,Padavi-2020-PRA, Moller-2020-NJP}.

%Unlike the disk BEC case, when we consider the evolution of 2D turbulence in a spherical BEC shell, any dipole configuration of the OV clusters is always associated with a finite angular momentum and therefore is prohibited if the BEC has zero angular momentum to begin with. In this situation, a novel quadrupole limiting configuration of the OV clusters with two pairs of like-signed clusters across two perpendicular diameters is expected (see Fig.~\ref{Fig1}~(b)). This configuration can induce a flow field with the highest kinetic energy while conserving the zero angular momentum.

In this Letter, we discuss our search for the exotic OV states in 2D spherical BECs. To our surprise, we find that OV clusters never form despite the annihilations of vortex pairs. We then present unequivocal analysis results to show that the spontaneous OV state in isolated BECs is not due to vortex-pair annihilations but instead is caused by vortices escaping the BEC boundary. Uncovering this true mechanism not only explains the absence of OV clusters in boundaryless 2D spherical BECs but also advances our knowledge of spontaneous vortex orders in 2D superfluid manifolds in general.

\textbf{Numerical method:} We model the dynamics of the BECs at low temperatures using the three-dimensional Gross-Pitaevskii equation (GPE)~\cite{Pitaevskii-2003-book}:
\begin{equation}
i\hbar \frac{\partial \psi}{\partial t}=\left[-\frac{\hbar^2}{2m}\nabla^2+U(\textbf{r},t)+g|\psi|^2\right]\psi,
\label{Eq1}
\end{equation}
where $\psi=|\psi|e^{i\phi}$ is the condensate wave function, $m$ is the particle mass, $g$ is the coupling constant, and $U$ is the external potential that confines the BEC. To generate quasi-2D BECs in both the disk and the spherical geometries for comparative studies, we adopt the confining potential used in Ref.~\cite{Simula-2014-PRL} to create a disk BEC:
\begin{equation}
U(\mathbf{r}) = U_0 \left[ \tanh \left( (r - R) / a_{osc} \right) + 1 \right] + \frac{1}{2} m \omega^2 z^2,
\end{equation}
where $U_0$ and $\omega$ are parameters pertinent to the trap strength in the radial plane and along the $z$-axis. $a_{osc}=\sqrt{{\hbar}/{m\omega}}$ is the characteristic trapping length in the $z$ direction that controls the disk thickness, and $R$ sets the disk radius. To create a spherical BEC shell, the following radial potential is used~\cite{Tononi-2019-PRL, Tononi-2020-PRL,Padavi-2020-PRA}:
\begin{equation}
U(\mathbf{r}) = \frac{1}{2} m \omega^2 \left(r - R \right)^2.
\end{equation}
For convenience, we normalize the time and length scales as $\tilde{t}=\omega t$ and $\tilde{r}=r/a_{osc}$ so the original GPE can be written in a dimensionless form:
\begin{equation}
i \frac{\partial \tilde{\Psi}}{\partial \tilde{t}} = \left[- \frac{1}{2}\tilde{\nabla}^2 + \frac{U}{\hbar \omega} + \tilde{g} |\tilde{\Psi}|^2\right] \tilde{\Psi}.
\label{Eq4}
\end{equation}
where $\tilde{\psi}=\psi/(\sqrt{N/a_{osc}^3})$ with $N$=$\int dV|\psi|^2$ being the total particle number. We select the trap parameters such that the normalized coupling constant $\tilde{g}=gN/\hbar\omega a_{osc}^3=\sqrt{125}\times10^4$ and $U_0/\hbar\omega=64$, matching those in Ref.~\cite{Simula-2014-PRL} and the experiment conducted by Neely \emph{et al.}~\cite{Neely-2013-PRL}. The radius for the disk BEC is set to $\tilde{R}=R/a_{osc}=30$ and for the spherical BEC shell is $\tilde{R}=15$ so the two BECs have the same surface areas.

We then numerically imprint~\cite{Padavi-2020-PRA,Kanai-2018-PRA,Kanai-2020-PRL} the velocity field of 80 vortices and 80 antivortices at random locations in the two BECs while keeping their angular momentum nearly zero~\cite{Simula-2014-PRL}. The Eq.~\ref{Eq4} is evolved in imaginary time for a short period so as to heal the vortex core structure~\cite{Chiofalo-2000-PRE}. The dynamical evolution of the condensate wavefunction is then obtained by numerically integrating Eq.~\ref{Eq4} with a spatial step of 0.1 and a time step of $10^{-3}$ using the forth-order Runge-Kutta method \cite{Press-1992-book}.
%$\triangle \tilde{x}$=$\triangle \tilde{y}$=$\triangle \tilde{z}$=, \triangle \tilde{t}$=2.5$\times

\textbf{Simulation results:} The evolution of the quasi-2D BEC from a typical initiate state in both the disk geometry and the spherical shell geometry can be seen in the movies in the Supplemental Material. In Fig.~\ref{Fig2}, we show snapshots of the condensate density on the $z=0$ plane for the disk BEC and on the $\tilde{r}=\tilde{R}$ surface for the spherical BEC shell. In the disk BEC, the like-signed vortices tend to form transient clusters that grow with time, which eventually lead to two counter-rotating persistent OV clusters. The annihilation of the vortices essentially ceases upon the formation of the OV clusters. These observations agree nicely with those of Ref.~\cite{Simula-2014-PRL}.

In the spherical BEC shell, the vortex-pair annihilations result in a somewhat more rapid decay of the total vortex number $N(\tilde{t})$, as shown in Fig.~\ref{Fig2}~(c) and (d). Note that in 2D BECs, two vortices can annihilate only via a multi-vortex interaction process~\cite{Nazarenko-2007-JLTP, Cidrim-2016-PRA,Baggaley-2018-PRA}. When a general $n$-vortex process controls the vortex decay, a scaling of $N(\tilde{t})\propto\tilde{t}^{-\frac{1}{n-1}}$ is expected~\cite{Baggaley-2018-PRA}. At large $\tilde{t}$ but before the OV clusters form in the disk BEC, we find that $N(\tilde{t})$ can be fitted well using this scaling with $n=2.4$ for the disk BEC and $n=3$ for the spherical shell BEC. The $n=3$ scaling is likely generic for pair annihilations in boundaryless quasi-2D BECs (see Supplemental Material). On the other hand, the $n=2.4$ scaling for the disk BEC indicates the presence of both two-vortex and three-vortex processes. Indeed, there are two distinct processes through which the vortices can decay in the disk BEC, i.e., pair annihilations and escaping from the disk boundary. The escaping process may be regarded as the annihilation of a vortex with its image charge in the presence of a second vortex, i.e., essentially a two-vortex process. According to Fig.~\ref{Fig2}~(c), about 1/3 of the decayed vortices in the disk BEC are caused by vortex escaping.

\begin{figure}[t]
\includegraphics[width=1.0\linewidth]{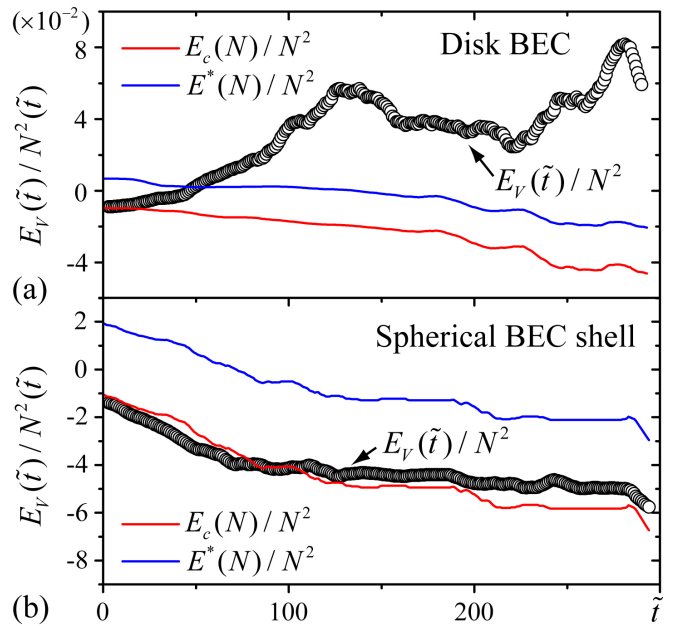}
\caption{Evolution of the incompressible kinetic energy $E_V$ associated with the vortices in a) the disk BEC and b) the spherical BEC shell. $E_c(N)$ is the threshold energy for transition to the negative temperature state, and $E^*(N)$ is a reference energy above which vortex clusters are readily observable.} \label{Fig3}
\end{figure}

Despite the more rapid annihilation of the vortex pairs in the spherical BEC shell, there appears to be no vortex clusters at any time (see Fig.~\ref{Fig2}~(b)). More concrete evidence showing whether or not OV clusters ever form in a BEC can be obtained from the evolution of the vortex energy. Note that the total kinetic energy of a BEC consists of three parts: an incompressible part due to the flow field induced by the vortices, a compressible part due to sound waves, and a quantum pressure term~\cite{Pethick-2008-book}. Many past studies evaluated the incompressible kinetic energy associated with the vortex system in planar BECs by first extracting the core locations of all vortices and then applying the following point-vortex Hamiltonian~\cite{Simula-2014-PRL,Billam-2014-PRL,Yu-2016-PRA,Reeves-2017-PRL, Groszek-2018-PRL}:
\begin{equation}
\resizebox{0.88\hsize}{!}{$
\begin{split}
{\cal H}=&-\frac{\rho_0\kappa^2}{4\pi}\bigg[\sum_{i<j} s_i s_j \ln(|\mathbf{r'_i}-\mathbf{r'_j}|^2 )-\sum_{i} s_i^2 \ln(1 - {r'_i}^2)\\
         &-\sum_{i < j} s_i s_j \ln \left(1 - 2\mathbf{r'_i}\cdot\mathbf{r'_j} + |r'_i|^2 |r'_j|^2 \right)\bigg],
\label{Eq:Disk}
\end{split}
$}
\end{equation}
where $\rho_0$ is the mean density, $\kappa=h/m$ is the quantized circulation, $\mathbf{r'_i}=\mathbf{r_i}/R$ denotes the normalized position vector of the $i$th vortex with a winding number $s_i=\pm1$. Here we adopt the same procedures. For vortices in the spherical shell, the following Hamiltonian is used~\cite{Bogomolov-1977-FD,Dritschel-2015-PRE}:
\begin{equation}
{\cal H}= - \frac{\rho_0\kappa^2}{4\pi}\sum_{i < j} s_i s_j \ln(1 - \mathbf{r'_i}\cdot\mathbf{r'_j}).
\label{Eq:Sph}
\end{equation}
The variations of the normalized incompressible kinetic energy $E_V=(4\pi/\rho_0\kappa^2){\cal H}$ in both BEC geometries are calculated and shown in Fig.~\ref{Fig3}. For reference purpose, we have also included in Fig.~\ref{Fig3} the threshold energy $E_c(N)$ above which a 2D neutral $N$-vortex system enters the negative temperature regime. This $E_c(N)$ is derived via a Markov chain Monte-Carlo method~\cite{Viecelli-1995-PoF} using the above Hamiltonians (see Supplemental Material). Since OV clusters appear only at energies significantly higher than $E_c(N)$~\cite{Yu-2016-PRA}, we also introduce a reference energy $E^*(N)$ at which the mean dipole (or quadrupole) moment of the vortices equals 30\% of the value for the limiting configuration depicted in Fig.~\ref{Fig1}. Above $E^*(N)$, clear vortex clusters are readily observable. Both $E_c(N)$ and $E^*(N)$ vary with $\tilde{t}$ as the total vortex number $N(\tilde{t})$ decays. From Fig.~\ref{Fig3}, one can see that for the disk BEC the vortex energy $E_V$ quickly rises to above $E^*(N)$, which explains why OV clusters were observed. On the contrary, $E_V$ for the spherical BEC shell barely gets above $E_c(N)$ and is always below $E^*(N)$, which thereby confirms that OV clusters never formed in the spherical BEC shell.

\begin{figure}[t]
\includegraphics[width=1.0\linewidth]{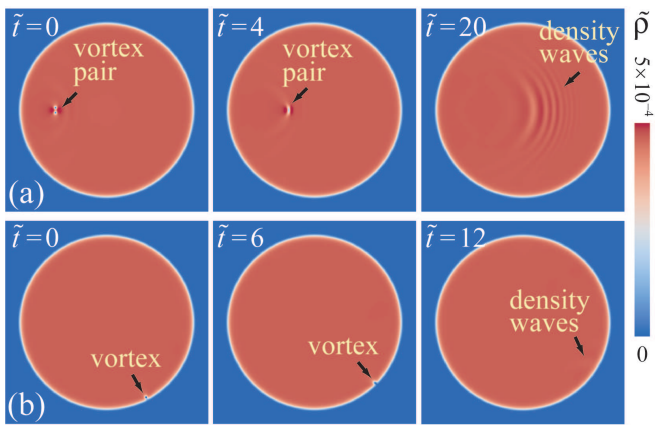}
\caption{GPE simulation showing density variations in the disk BEC when: a) a vortex-antivortex pair undergoes annihilation; and b) a vortex merges into the disk boundary.
} \label{Fig4}
\end{figure}

The contrasting fate of the vortices in the disk BEC and the spherical BEC shell calls for an explanation. As we discussed earlier, the vortices in the spherical BEC shell can decay only via pair annihilations, whereas in the disk BEC they can decay via both pair annihilations and escaping from the boundary. To better understand the consequence of this difference, we simulated the annihilation of an isolated vortex pair and the escaping of a single vortex in the disk BEC using GPE. For the annihilation test, we first prepare a vortex-antivortex pair at close separation and then evolve Eq.~\ref{Eq4} with a small added damping, similar to that discussed in Ref.~\cite{Baggaley-2018-PRA}, so the two vortices approach each other while the pair propagates. When the vortex separation is about the core size, we set $\tilde{t}=0$ and remove the added dissipation so the subsequent annihilation process is not affected by artificial damping. Similar procedures are adopted for the single vortex near the disk boundary. The results are shown in Fig.~\ref{Fig4}. One can see clearly that the pair annihilation in bulk BEC generates intense sound waves so as to conserve the momentum carried by the pair. On the contrary, in the vortex escaping process, the vortex merges into the zero-density region, which hardly generates any sound waves.

%One can see clearly that the pair annihilation and the collapsing of the vortex cores in bulk BEC lead to intense sound-wave emission. On the contrary, in the vortex escaping process, the vortex core merges into the zero-density region, which hardly generates any sound waves.

Note that the sound waves in the BEC can damp out the vortex motion and dissipate the incompressible kinetic energy possessed by the vortex system~\cite{Nazarenko-2007-JLTP}. This process is similar in nature to the mutual friction damping on quantized vortices in superfluid helium caused by the normal-fluid component~\cite{Gao-2016-PRB,Gao-2018-PRB,Yui-2020-PRL}. Therefore, one may draw the following conclusions: 1) the pair annihilation process alone does not lead to the formation of OV clusters due to the intense sound emission; and 2) the escaping of the vortices from the BEC boundary, which increases the mean energy of the vortices with minimal sound emission, is the true mechanism responsible for spontaneous vortex orders. To verify these conclusions, we present two complementary tests that can produce unequivocal supporting evidences.

\begin{figure}[b]
\includegraphics[width=1.0\linewidth]{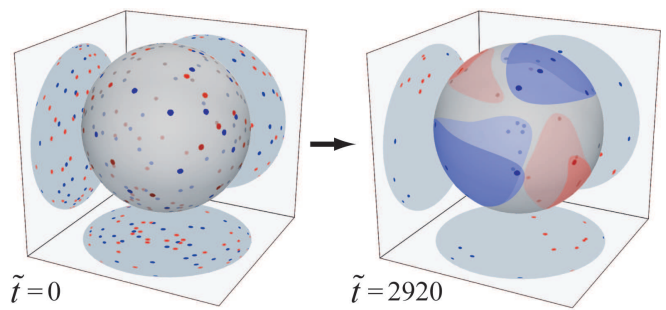}
\caption{Point-vortex model simulation of the vortex dynamics on 2D spherical surface from the same initial state as in our GPE simulation.
} \label{Fig5}
\end{figure}

\textbf{Complementary tests:} In the first test, we examine the ideal dynamics of the vortices on the spherical surface ($\tilde{R}=15$) without sound waves. To this do, we consider point vortices with the same initial distribution as in our GPE simulation and evolve them using the following equation of motion derived from the Hamiltonian in Eq.~(\ref{Eq:Sph})~\cite{Bogomolov-1977-FD,Dritschel-2015-PRE}:
\begin{equation}
\frac{d\mathbf{r'_i}}{d\tilde{t}}=\frac{1}{2\tilde{R}^2}\sum_{j\neq i}\frac{\mathbf{r'_j}\times\mathbf{r'_i}}{1-\mathbf{r'_j}\cdot\mathbf{r'_i}}.
\end{equation}
To mimic the vortex-pair annihilation process in GPE, we remove vortex-antivortex pairs whenever the arc-length separation between two vortices is less than $0.03\tilde{R}$~\cite{Simula-2014-PRL}. At large $\tilde{t}$, we find that four vortex clusters form spontaneously as shown in Fig.~\ref{Fig5}, which eventually evolve towards the limiting configuration given in Fig.~\ref{Fig1}~(b). This dynamics is not surprising, because removing a vortex pair at close separation essentially amounts to subtracting a large negative quantity from the Hamiltonian. Therefore, the energy of the point-vortex system steadily increases with time, which inevitably leads to the formation of OV clusters. The exact time it takes before OV clusters emerge depends on the threshold separation for vortex-pair removal. This test shows that the pair-annihilation based evaporative-heating mechanism would work only in the absence of sound waves. Our result also calls for caution in using the point-vortex model to understand the vortex dynamics in real BECs.

\begin{figure}[t]
\includegraphics[width=1.0\linewidth]{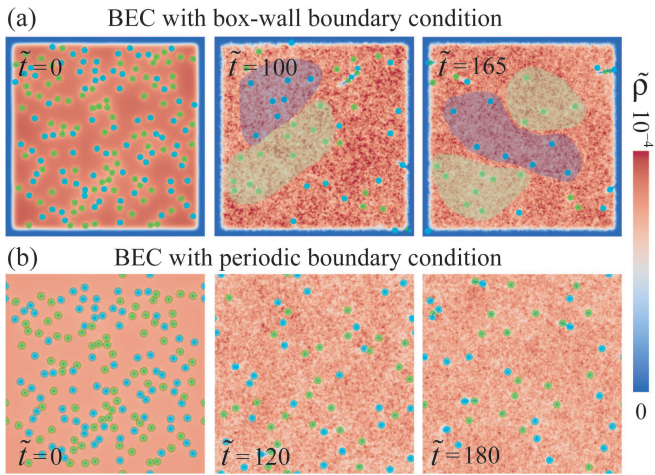}
\caption{GPE simulation of the vortex dynamics in quasi-2D square BEC with a) box-wall boundary condition; and b) periodic boundary condition.
} \label{Fig6}
\end{figure}

In the second test, we conduct a GPE simulation with 80 vortices and 80 antivortices at random locations in a square-shaped planar quasi-2D BEC. We adopt the same trapping parameters $U_0$ and $\omega$ as for the disk BEC and set the side length of the square to $\tilde{R}=50$ so its area is also similar. The advantage of the square shape is that we can now easily change the box-wall boundary condition (i.e., with the hyperbolic tangent potential) to a periodic boundary condition~\cite{Baggaley-2018-PRA} so that the vortex dynamics in the same BEC geometry with and without the vortex-escaping mechanism can be compared directly. Fig.~\ref{Fig6} shows representative snapshots of the BEC density from the same initial state with the two different boundary conditions. Again, clear OV clusters are seen only in the case with the box-wall boundary, which unambiguously verifies that the emergent vortex order is caused by vortex escaping from the BEC boundary.

In summary, we have examined the evolution of vortices in both planar and spherical 2D BECs. A comprehensive understanding of the mechanism underlying the spontaneous vortex orders is achieved, which represents a major progress in the study of the far-from-equilibrium dynamics of 2D superfluids. Our findings may also motivate future experiments in 2D spherical BECs at the international space station.

\begin{acknowledgments}
The authors thank S. Nazarenko and K. Helmerson for stimulating discussions. The authors also acknowledge the support by the National Science Foundation under Grant No. DMR-2100790. The work was conducted at the National High Magnetic Field Laboratory, which is supported by National Science Foundation Cooperative Agreement No. DMR-1644779 and the state of Florida.
\end{acknowledgments}

\bibliography{2D-QT}
\end{document}

\title{Supplemental Materials for: The true mechanism of spontaneous order from turbulence in two-dimensional superfluid manifolds}
%Angular Momentum Transport in Bose-Einstein Condensates by Solitonic ``Corkscrew''
%Torque and angular momentum transfer by nontopological phase defects in Bose-Einstein Condensates
\author{Toshiaki Kanai}
\affiliation{National High Magnetic Field Laboratory, 1800 East Paul Dirac Drive, Tallahassee, Florida 32310, USA}
\affiliation{Department of Physics, Florida State University, Tallahassee, Florida 32306, USA}

\author{Wei Guo}
%\email[Corresponding: ]{wguo@magnet.fsu.edu}
\affiliation{National High Magnetic Field Laboratory, 1800 East Paul Dirac Drive, Tallahassee, Florida 32310, USA}
\affiliation{Mechanical Engineering Department, FAMU-FSU College of Engineering, Florida State University, Tallahassee, Florida 32310, USA}

\maketitle
\section{Point-vortex thermodynamics}
In the framework of the Gross-Pitaevskii equation (GPE), the total kinetic energy of a BEC can be decomposed into three parts: an incompressible part due to the flow field induced by the vortices, a compressible part due to sound waves, and a quantum pressure term~\cite{Pethick-2008-book}. To evaluate the incompressible kinetic energy associated with the vortex system, a commonly adopted method is to extract the core locations of the vortices and then calculate this energy using a point-vortex Hamiltonian~\cite{Simula-2014-PRL,Billam-2014-PRL,Yu-2016-PRA,Reeves-2017-PRL, Groszek-2018-PRL}. For a planar BEC, this Hamiltonian is:
\begin{equation}
\resizebox{0.88\hsize}{!}{$
\begin{split}
{\cal H}=&-\frac{\rho_0\kappa^2}{4\pi}\bigg[\sum_{i<j} s_i s_j \ln(|\mathbf{r'_i}-\mathbf{r'_j}|^2 )-\sum_{i} s_i^2 \ln(1 - {r'_i}^2)\\
         &-\sum_{i < j} s_i s_j \ln \left(1 - 2\mathbf{r'_i}\cdot\mathbf{r'_j} + |r'_i|^2 |r'_j|^2 \right)\bigg],
\label{Eq:Disk}
\end{split}
$}
\end{equation}
where $\rho_0$ is the mean density of the BEC, $\kappa=h/m$ is the quantized circulation, $\mathbf{r'_i}=\mathbf{r_i}/R$ is the normalized position vector of the $i$th vortex with a winding number $s_i=\pm1$. For vortices in a spherical BEC shell, the corresponding point-vortex Hamiltonian is~\cite{Bogomolov-1977-FD,Dritschel-2015-PRE}:
\begin{equation}
{\cal H}= - \frac{\rho_0\kappa^2}{4\pi}\sum_{i < j} s_i s_j \ln(1 - \mathbf{r'_i}\cdot\mathbf{r'_j}).
\label{Eq:Sph}
\end{equation}
When the normalized vortex energy $E=(4\pi/\rho_0\kappa^2){\cal H}$ is higher than a threshold $E_c$, the vortex system enters the negative temperature regime. At sufficiently high energies, Onsager vortex (OV) clusters can emerge. In order to determine these thermodynamic energy levels for reference purpose, a Markov chain Monte-Carlo method can be adopted~\cite{Viecelli-1995-PoF}. The relevant procedures have been discussed in detail for planar disk BECs~\cite{Yu-2016-PRA}. Here we outline the major steps for the 2D spherical BEC case.

\begin{figure}[t]
\includegraphics[width=1.0\linewidth]{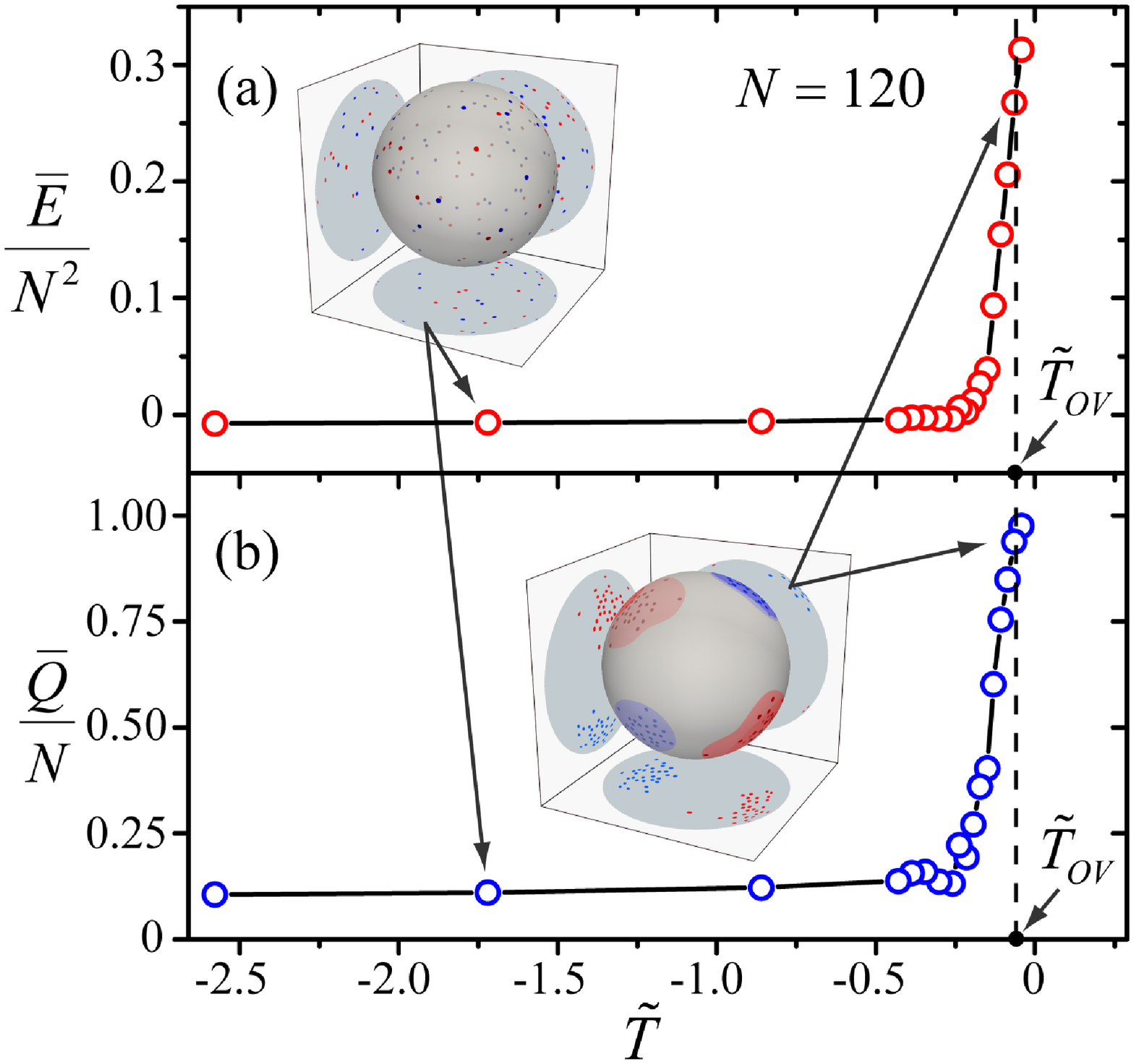}
\caption{Variations of a) the mean energy $\bar{E}$ and b) the mean quadrupole moment $\bar{Q}$ as a function of the normalized temperature $\tilde{T}$ for a neutral point-vortex system in a spherical shell with $N=120$ vortices and zero BEC angular momentum. $\tilde{T}_{OV}=-1/16$ denotes the ideal point-vortex super-condensation transition temperature.
} \label{Fig1-App}
\end{figure}

We consider a neutral point-vortex system with a total vortex number $N$ in a spherical BEC shell ($\tilde{R}=R/a_{osc}=15$) having zero angular momentum. To evaluate the thermodynamic properties of this vortex system, a large ensemble (i.e., $5\times10^6$) of vortex configurations for a given temperature $T$ are generated based on the Boltzmann distribution $e^{-E/N\tilde{T}}$ using the Monte Carlo method as detailed in Ref.~\cite{Viecelli-1995-PoF}, where $\tilde{T}=T/T_{0}$ is the normalized temperature with $T_0=N\rho_0\kappa^2/4\pi k_B$. We restrict the generated vortex configurations to have negligible vortex dipole moment $\mathbf{d}=\sum_is_i\mathbf{r'_i}$ and therefore nearly zero BEC angular momentum. The mean energy of the vortex system $\bar{E}(\tilde{T})$ is obtained as the average of $E$ over all vortex configurations. In Fig.~\ref{Fig1-App}~(a), we plot $\bar{E}$ versus $\tilde{T}$ for a representative vortex system with $N=120$. Besides the vortex energy, we have also calculated the quadrupole moment $Q$ for each vortex configuration, defined as $Q=(\sum_{l}q^2_l)^{1/2}$ where $q_l$ ($l=x,y,z$) is the eigenvalue of the following quadrupole tensor:
\begin{equation}
Q_{ll'}=\frac{1}{2}\sum_is_i[3(\mathbf{r'_i}\cdot \hat{\mathrm{e}}_l)(\mathbf{r'_i}\cdot \hat{\mathrm{e}}_{l'})-\delta_{ll'}].
\end{equation}
The maximum quadrupole moment $Q_{Max}/N\simeq3\sqrt{2}/4$ is achieved in the limiting vortex configuration as shown in Fig. 1 (b) in the paper, where the vortices form four compact clusters, each containing $N/4$ like-signed vortices. The mean quadrupole moment $\bar{Q}(\tilde{T})$ at different $\tilde{T}$ is determined as the ensemble average of $Q$ and is shown in Fig.~\ref{Fig1-App}~(b). As the temperature approaches $-0$, both $\bar{E}(\tilde{T})$ and $\bar{Q}(\tilde{T})$ rise sharply, signifying a transition to the Onsager-vortex phase. Indeed, through an energy-entropy balancing analysis~\cite{Kraichnan-1980-RPP}, one can derive a temperature $\tilde{T}_{OV}$ above which the vortex system would undergo an super-condensation transition. The obtained $\tilde{T}_{OV}$ for the disk BEC is -1/4~\cite{Simula-2014-PRL,Billam-2014-PRL,Yu-2016-PRA}, and a similar analysis gives $\tilde{T}_{OV}=-1/16$ for the spherical BEC shell. In Fig.~\ref{Fig1-App}, we also include two representative microcanonical vortex configurations at temperatures below and close to $\tilde{T}_{OV}$.

\begin{figure}[t]
\includegraphics[width=1.0\linewidth]{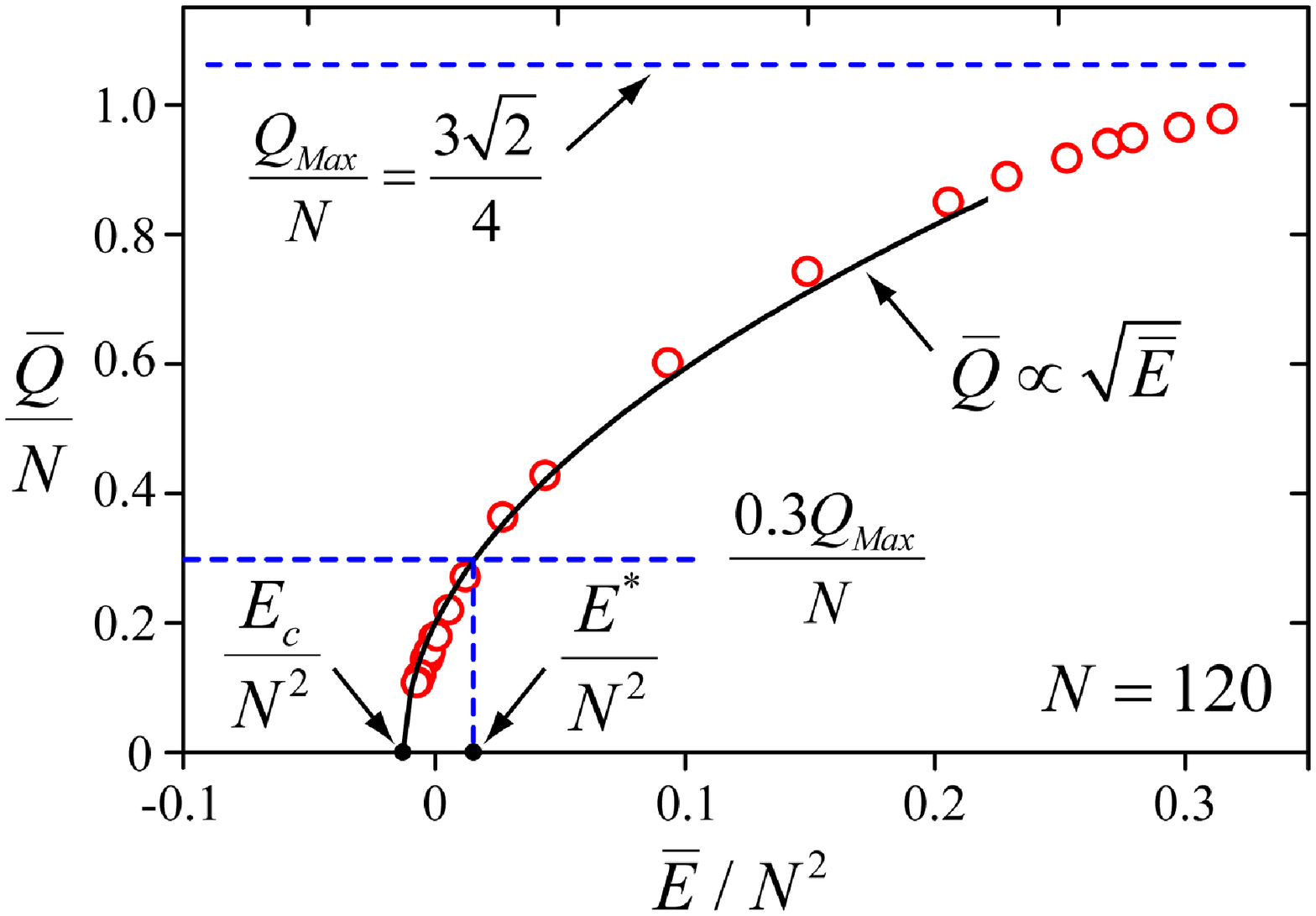}
\caption{The derived mean quadrupole moment $\bar{Q}$ as a function of the mean vortex energy $\bar{E}$ for the vortex system considered in Fig.~\ref{Fig1-App}. The reference energy levels $E_c$ and $E^*$ introduced in the text can be determined.
} \label{Fig2-App}
\end{figure}

Now we can proceed to evaluate some key reference energies. The threshold energy $E_c$ is essentially the value of $\bar{E}$ as $\tilde{T}$ approaches $-\infty$. To determine $E_c$ reliably, we follow the method as discussed in Ref.~\cite{Yu-2016-PRA} and plot $\bar{Q}(\tilde{T})$ versus $\bar{E}(\tilde{T})$ in Fig.~\ref{Fig2-App}. The data near $\bar{Q}=0$ follows a $\sqrt{\bar{E}}$ scaling~\cite{Yu-2016-PRA}. $E_c$ can be determined as the intersect of this scaling curve with the $\bar{E}$-axis. As for the energy associated with the formation of OV clusters, we introduce a phenomenological reference energy $E^*$ at which the mean quadrupole moment equals 30\% of $Q_{Max}$, instead of using the energy corresponding to $\tilde{T}_{OV}$. This is because $\tilde{T}_{OV}$ refers to the idealized super-condensation phase transition, but OV clusters can emerge even at slightly lower $\tilde{T}$. Comparing Fig.~\ref{Fig1-App} and Fig.~\ref{Fig2-App}, one can see that the $E^*$ we introduced is at the level where the $\bar{E}$ curve starts to rise sharply.

\begin{figure}[t]
\includegraphics[width=1.0\linewidth]{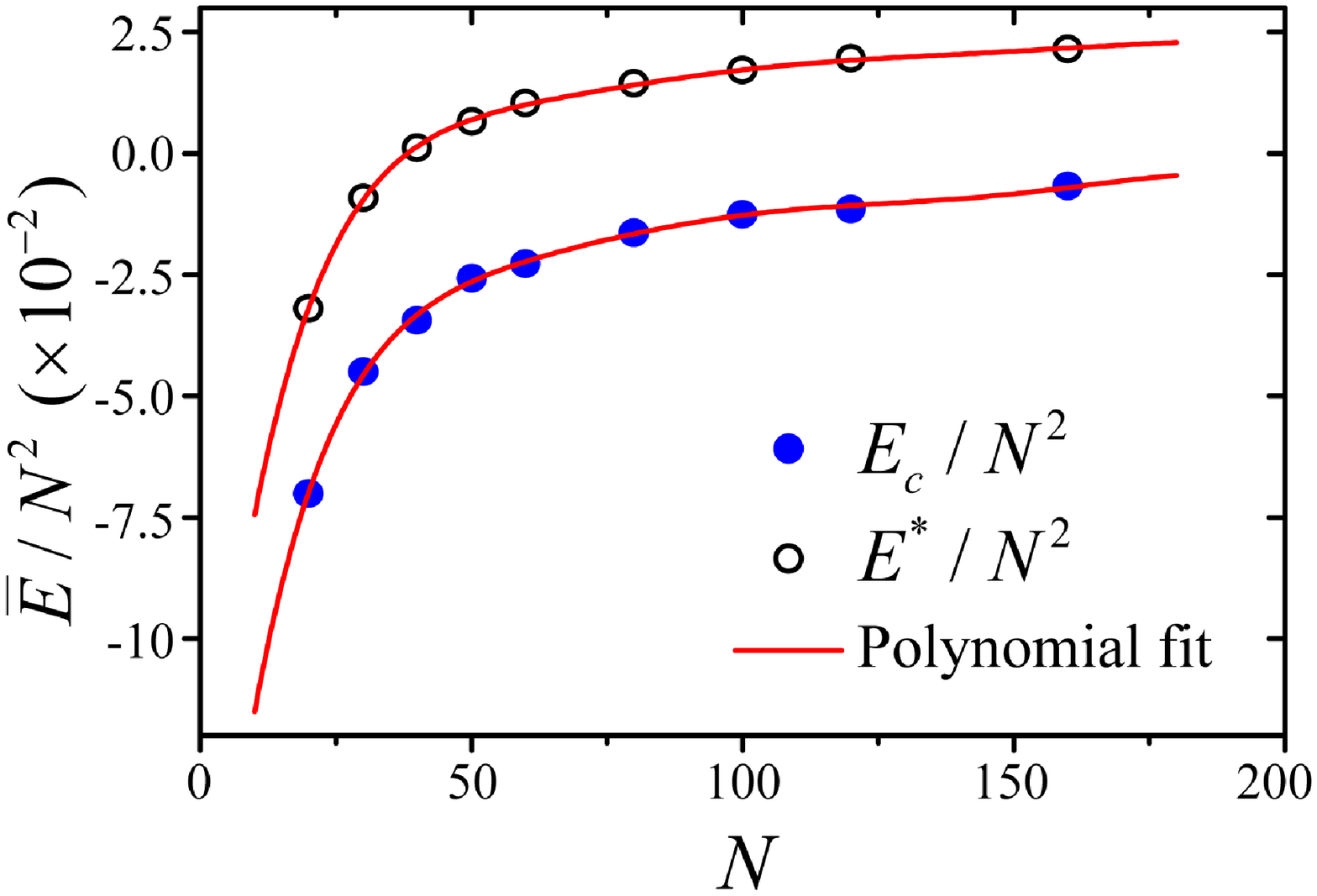}
\caption{Variations of the derived $E_c$ and $E^*$ with the total vortex number $N$. The solid red curves represent polynomial fits to the data with the form $E=\sum_{i=0}^7a_iN^i$.
} \label{Fig3-App}
\end{figure}

We then repeat the above analysis for vortex systems with different $N$. The obtained $E_c$ and $E^*$ are collected in Fig.~\ref{Fig3-App}. In order for convenient comparison with the incompressible kinetic energy $E_V(\tilde{t})$ we calculated for the vortex system in our GPE simulation, a polynomial fits of the form $E=\sum_{i=0}^7a_iN^i$ is performed to both the $E_c$ data and the $E^*$ data so that their dependance on the vortex number $N$ can be determined. In Fig. 3 in the paper, at a given $\tilde{t}$, the vortex number $N(\tilde{t})$ is known. We can then include the corresponding $E_c(N(\tilde{t}))$ and $E^*(N(\tilde{t}))$ to the figure.

\section{Decay scaling of vortex number}
In an ideal 2D BEC without any added damping, a vortex and an antivortex alone cannot decay via pair annihilation. Instead, they would form a stable pair and travel at a constant velocity~\cite{Donnelly-book}. The interaction with a third vortex is needed in order to dissipate the energy of the vortex pair so that the two vortices can approach each other and annihilate. This annihilation then leads to the generation of a long-lived nonlinear density wave, which was first identified by Nazarenko and Onorato as a soliton~\cite{Nazarenko-2007-JLTP} and was later denoted as the ``crescent-shaped'' wave by Kwon \emph{et al.}~\cite{Kwon-2014-PRA} and the ``vortexonium'' by Groszek \emph{et al.}~\cite{Groszek-2016-PRA}. This nonlinear wave may collide with a fourth vortex and eventually decay into phonons~\cite{Groszek-2016-PRA,Cidrim-2016-PRA}. Therefore, in a boundaryless ideal 2D BEC, the vortices are expected to decay via a four-vortex interaction process, which was confirmed by Baggaley and Barenghi in their study of decaying homogeneous turbulence in an ideal 2D square BEC with a periodic boundary condition~\cite{Baggaley-2018-PRA}. These authors found that at large decay times the total vortex number scales as $N(\tilde{t})\propto\tilde{t}^{-\frac{1}{3}}$. Note that when a general $n$-vortex process controls the vortex decay, a scaling of $N(\tilde{t})\propto\tilde{t}^{-\frac{1}{n-1}}$ is expected. Therefore, their result suggests that $n=4$. Nonetheless, they also showed that when some damping was intentionally added to the 2D BEC, the decay of $N(\tilde{t})$ can change to a scaling with $n=3$. This is because the added damping can dissipate the soliton wave without the need for a fourth vortex. At late times when the vortex number $N$ is very small, the dissipation in the BEC can bring the vortex and the closest antivortex together for annihilation without the presence of other vortices, which could eventually lead to a two-vortex decay scaling~\cite{Baggaley-2018-PRA}.

In our paper, we showed in Fig. 2 that the decay of the total vortex number $N(\tilde{t})$ exhibts the scaling of $n=2.4$ for the quasi-2D disk BEC and $n=3$ for the spherical BEC shell. In a qusi-2D BEC with a finite thickness, the interaction between the sound waves and the vortices is strong as compared to that in ideal zero-thickness 2D BECs~\cite{Simula-2014-PRL}. This enhanced interaction likely plays the role of the added damping as in ideal 2D BECs, which therefore could result in the observed $n=3$ decay scaling of $N(\tilde{t})$ in the boundaryless spherical BEC shell. Based on this hypothesis, the $n=2.4$ decay scaling found in the quasi-2D disk BEC can be interpreted naturally as due to the interplay of vortex-pair annihilations (i.e., a three-vortex process) and vortices escaping from the disk boundary (i.e., a two-vortex process as discussed in the paper).

\begin{figure}[t]
\includegraphics[width=1.0\linewidth]{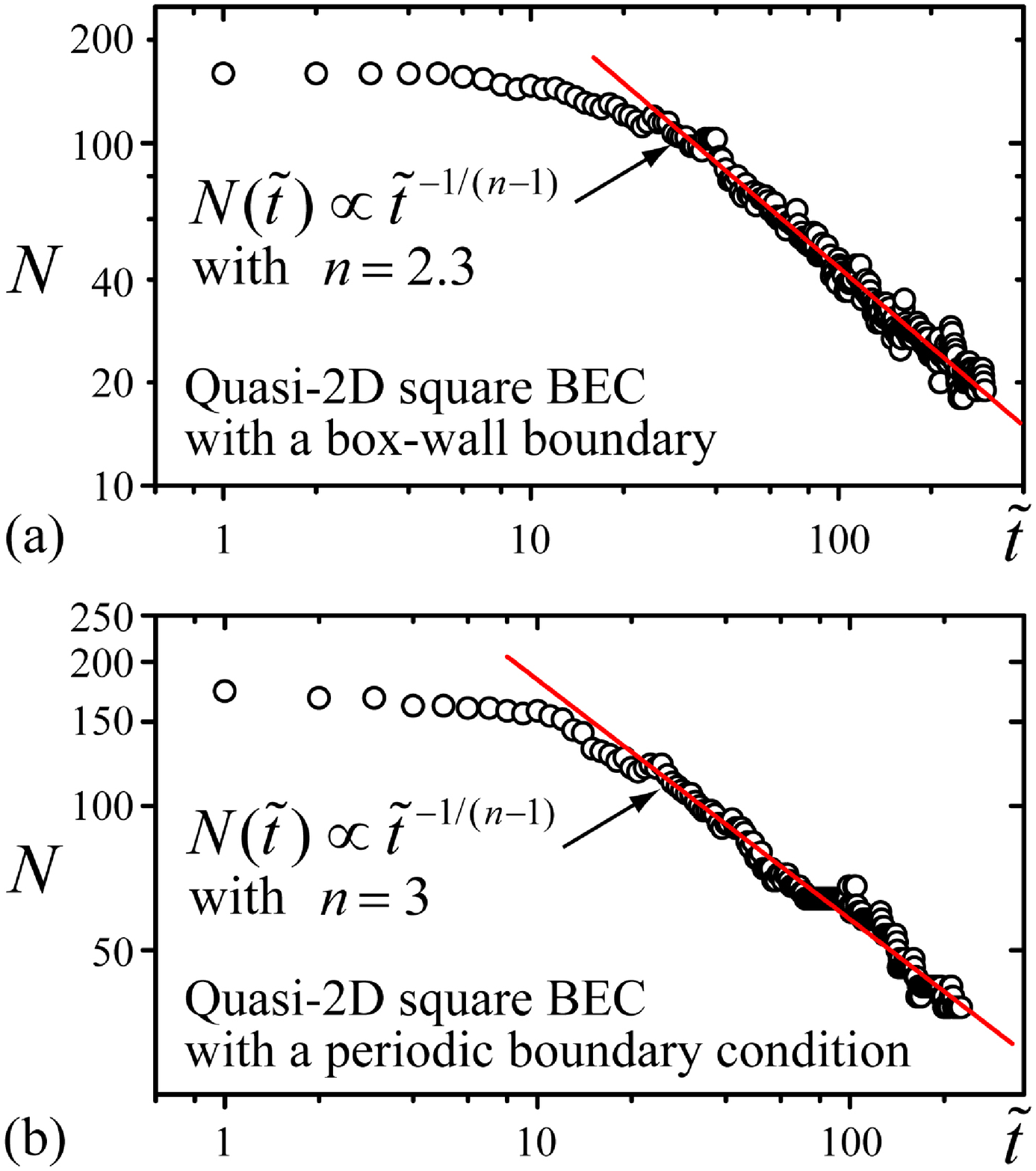}
\caption{GPE simulation of the time evolution of the total vortex number $N(\tilde{t})$ in a square BEC with a) a box-wall boundary condition; and b) a periodic boundary condition. These results are pertinent to the cases shown in Fig. 6 in the paper.
} \label{Fig4-App}
\end{figure}

To support this view, we have examined the decay of the vortex number in quasi-2D square BECs with both the box-wall boundary condition (i.e., with the hyperbolic tangent potential as described in the paper) and the periodic boundary condition. The variations of the vortex number $N(\tilde{t})$ pertinent to the two cases presented in Fig. 6 in the paper are shown in Fig.~\ref{Fig4-App}. For the case with the periodic boundary condition, we again observed the $n=3$ decay scaling at late times, which therefore supports the generic nature of this scaling for vortex-number decay in boundaryless quasi-2D BECs. For the case with the box-wall boundary, a decay scaling of $n=2.3$ is observed, which is close to that in the disk BEC bounded by the same type boundary. We would like to add that we have also examined the variation of $N(\tilde{t})$ in an ideal 2D square BEC with the periodic boundary condition and confirmed the $n=4$ decay scaling as reported by Baggaley and Barenghi~\cite{Baggaley-2018-PRA}. Therefore, these observations together support our view that as the BEC thickness increases from zero to finite, the enhanced sound-vortex interaction can alter the vortex-number decay scalings.

\section{Supplemental Movies}
\textbf{Movie S1:} Evolution of the quasi-2D disk BEC, simulated using the Gross-Pitaevskii model. The BEC is initially imprinted with 80 vortices and 80 antivortices and has nearly zero angular momentum with respect to the disk center, as described in text. The movie shows the evolution of the condensate density on the $z=0$ plane. The locations of the vortices and antivortices are marked with blue and green dots, respectively.

\textbf{Movie S2:} Evolution of the quasi-2D spherical shell BEC, simulated using the Gross-Pitaevskii model. The BEC is initially imprinted with 80 vortices and 80 antivortices and has nearly zero angular momentum. The movie shows the evolution of the condensate density on the $\tilde{r}=\tilde{R}$ surface. The locations of the vortices and antivortices are marked with blue and green dots, respectively.

\bibliography{Suppl-reference}